\documentclass[11pt,german,a4]{article}
\usepackage{ifpdf}
\ifpdf
  \usepackage[pdftex]{graphicx}
\else
  \usepackage[dvips]{graphicx}
\fi
\usepackage[sort]{natbib}

\bibpunct{(}{)}{;}{a}{}{,}

\usepackage{url}
\urldef{\makro}{\url}{http://p3d.sourceforge.net}

\author{
Martin M. Roth,
Christer Sandin
\\
\small innoFSPEC Center for Innovation Competence \\
\small at Astrophysikalisches Institut Potsdam\\
\small An der Sternwarte 16, D-14482 Potsdam, Germany\\
\texttt{mmroth@aip.de}
}

\title{Visualization of Data from Integral Field\\ 
Spectroscopy and the P3d Tool}

\begin{document}
\maketitle
\thispagestyle{empty} 

\begin{abstract}
\noindent Integral Field Spectroscopy is a powerful observing technique
for Astronomy that is becoming available at most ground-based observatories 
as well as in space. The complex data obtained with this technique require
new approaches for visualization. Typical requirements and the p3d tool,
as an example, are discussed.
\end{abstract}

\section{Integral Field Spectroscopy}

\begin{figure}[h!]
\centering
\includegraphics[width=8cm]{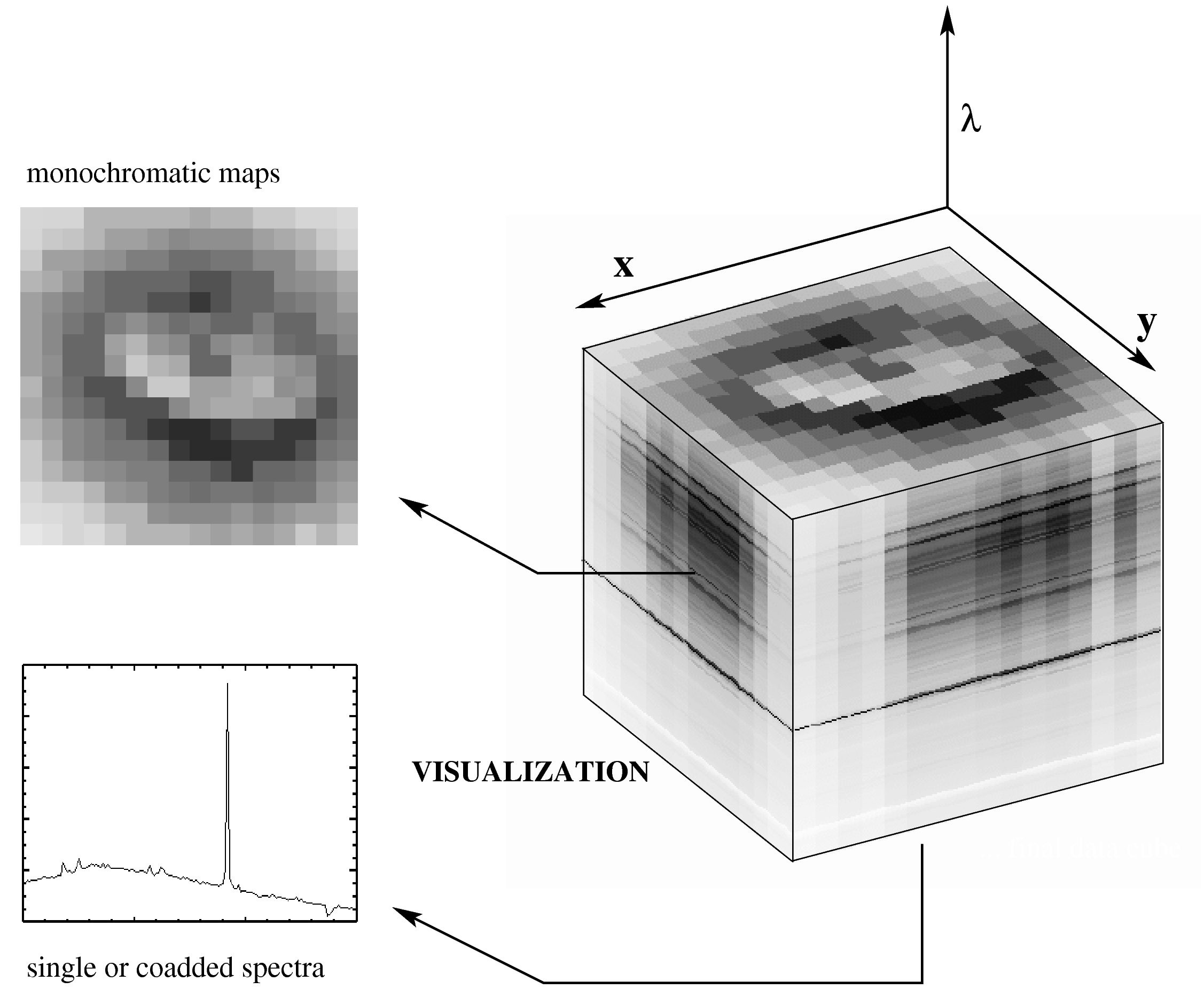}
\caption{Three-dimensional dataset as the result from IFS (there are two
  spatial coordinates and one wavelength coordinate). The datacube can be
  visualized as a stack of quasi-monochromatic images, or, alternatively,
  as an assembly of $n\times m$ spectra.}
\label{DATACUBE}
\end{figure}

Integral Field Spectroscopy (IFS), which is -- somewhat confusingly -- 
also called ``3D'' or ``tri-dimensional'' spectroscopy'', ``two-dimensional'' 
or ``area'' spectroscopy and so forth, and which is known in other areas beyond 
astronomy under terms like e.g.\ ``hyperspectral imaging'', is a powerful
observing technique that has been introduced and refined over the past
two decades. It is now at the verge of becoming a standard tool, which is
available at most modern telescopes \citet{Ro:10}.
For practical reasons, some users have, furthermore, adopted the intuitively descriptive
terminology ``3D'' as a reference to the datacube, which is thought to be the
product of an observation (cf.\ Fig.~\ref{DATACUBE}). 

IFS is an astronomical observing method that in a single exposure creates
spectra of (typically many) spatial elements (``spaxels'') simultaneously 
over a two-dimensional field-of-view (FoV) on the sky. Owing to this sampling 
method, each spaxel can be associated with its individual spectrum. Once all 
of the spectra have been extracted from the detector frame, in the 
data-reduction process, it is possible to reconstruct maps at arbitrary
wavelengths. For instruments with an orthonormal spatial sampling geometry, 
the spectra can be arranged
on the computer to form a three-dimensional array, which is most commonly
called a ``\emph{datacube}''. Datacubes are also well-known as the natural
data product in radio astronomy. However, there are many integral
field spectrographs which do not sample the sky in an orthonormal
system. In this case the term datacube is misleading. Also, atmospheric
effects, in particular in the optical wavelength regime, make the term 
inappropriate in the most general case. 

Instruments that create three-dimensional datasets in the above
mentioned sense, however not simultaneously but rather in some
process of sequential data acquisition (scanning) -- e.g.\ tunable filter
(Fabry-Perot) instruments, scanning long-slits, etc.\ -- are not strictly 
3D spectrographs according to this definition.

\subsubsection*{Image Dissection, Spatial Sampling, Spectra}

\begin{figure}[t!]
\centering
\includegraphics[width=12cm,angle=0]{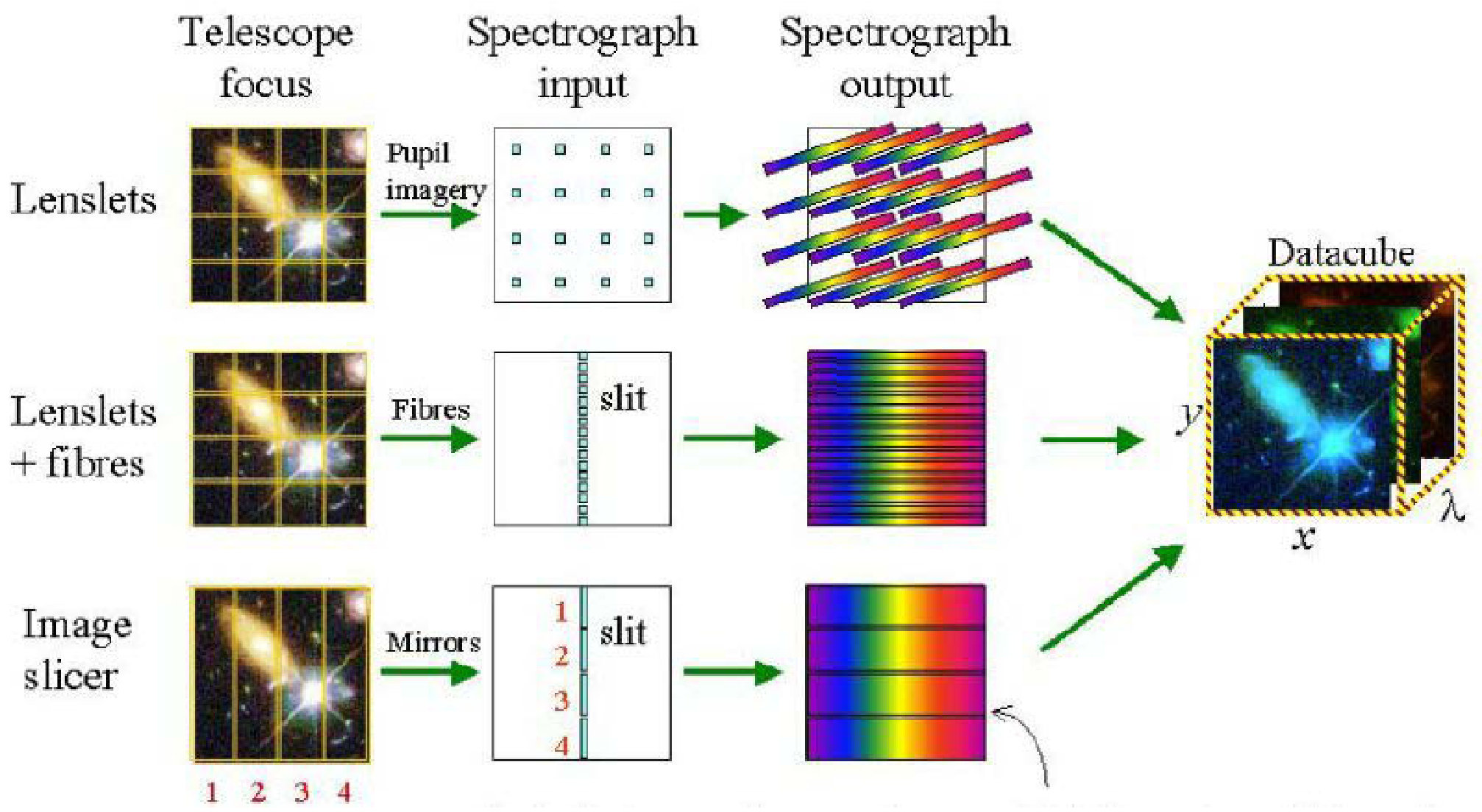}
\caption{The three major principles of operation of
present-day IFUs (source: J.Allington-Smith).}
\label{IFUTYPES}
\end{figure}

Integral field spectrographs have been built based on different methods of 
dissecting the FoV into spaxels, e.g.\ optical fiber bundles,
lens arrays, optical fibers coupled to lens arrays, or slicers
(Fig.~\ref{IFUTYPES}).

\pagebreak

\begin{figure}[h!]
\centering
\includegraphics[width=4cm,angle=0]{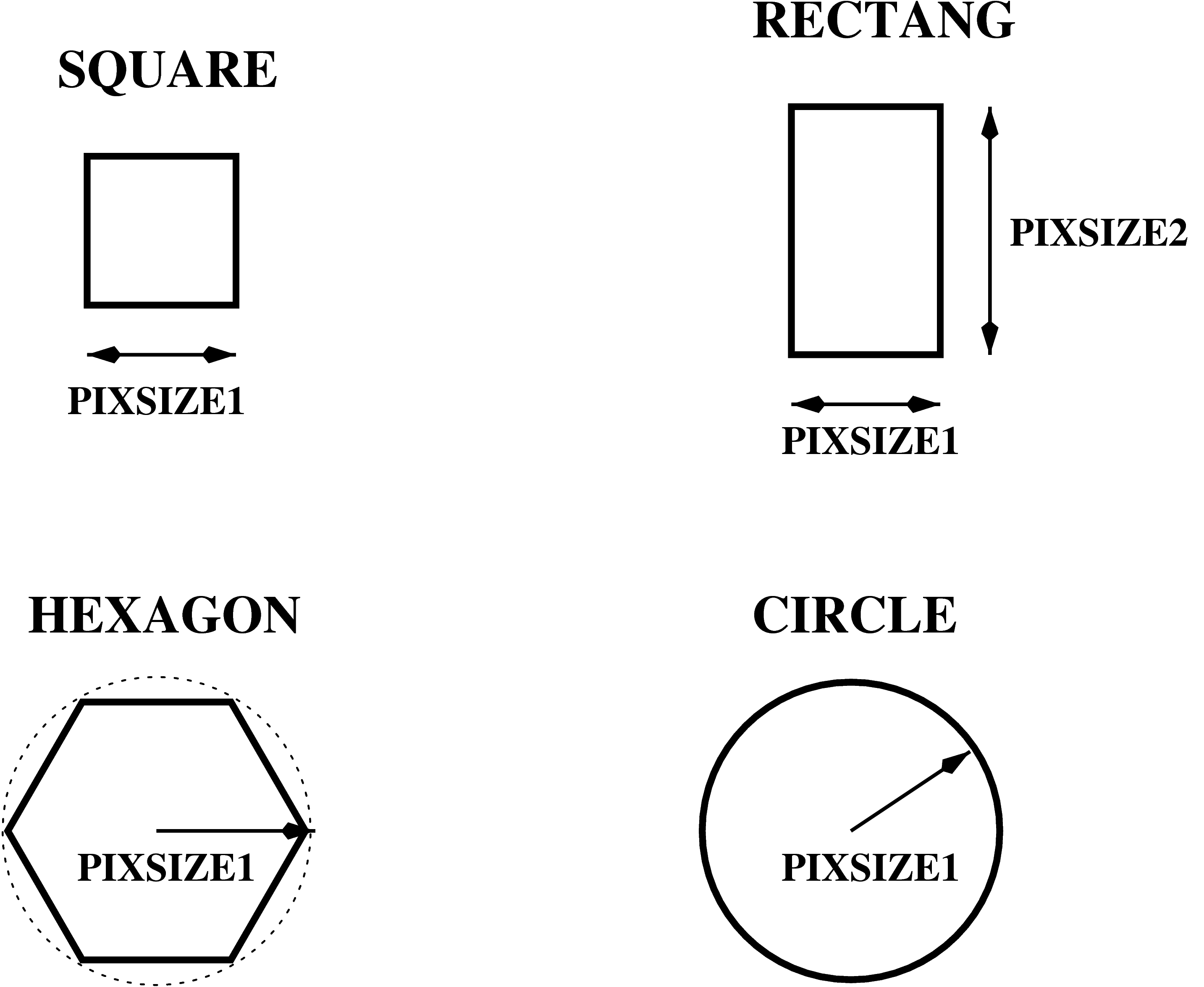}
\caption{Four different types of spaxel geometries: square,
rectangular, hexagonal, and bare fiber (circular).}
\label{SPAXELS}
\end{figure}

The term \emph{spaxel} was introduced  in order to
distinguish spatial elements in the image plane of the \emph{telescope}
from \emph{pixels}, which are the spatial elements in the image plane of
the \emph{detector} \citep{Ki:04}. The optical 
elements that accomplish the sampling of the sky are often called 
``integral field units'' (IFUs), and IFS is also sometimes called
``IFU spectroscopy''. Spaxels can have different shapes and sizes, depending 
on instrumental details and the type of IFU (Fig.~\ref{SPAXELS}).

Contrary to the persuasive implication of the datacube picture, 
IFUs do not necessarily sample the sky on a regular grid: 
e.g.\ fiber bundles, where, due to the manufacturing process,
individual fibers cannot be arranged to arbitrary precision.
Even if the manufacture of an IFU allows to create a
perfectly regular sampling pattern, e.g.\ in the case of a hexagonal
lens array, the sampling is not necessarily orthonormal. Moreover,
real optical systems create aberrations and, sometimes, non-negligible
field distortions, in which case the spectra extracted from the detector 
do not sample an orthogonal FoV on the sky. 
Furthermore, the sampling method may be contiguous
(e.g.\ lens array) fith a fill factor very close to unity, or
non-contiguous (e.g.\ fiber bundle) with a fill factor of significantly
less than 100\%. In all of these cases, it is possible to reconstruct
maps at a given wavelength through some process of interpolation and,
repeating this procedure over all wavelengths, to convert the result 
into a datacube. Note, however, that interpolation often produces
artifacts and generally involves loss of information.

\begin{figure}[h!]
\begin{minipage}[b]{6cm}
\includegraphics[width=6cm,angle=0]{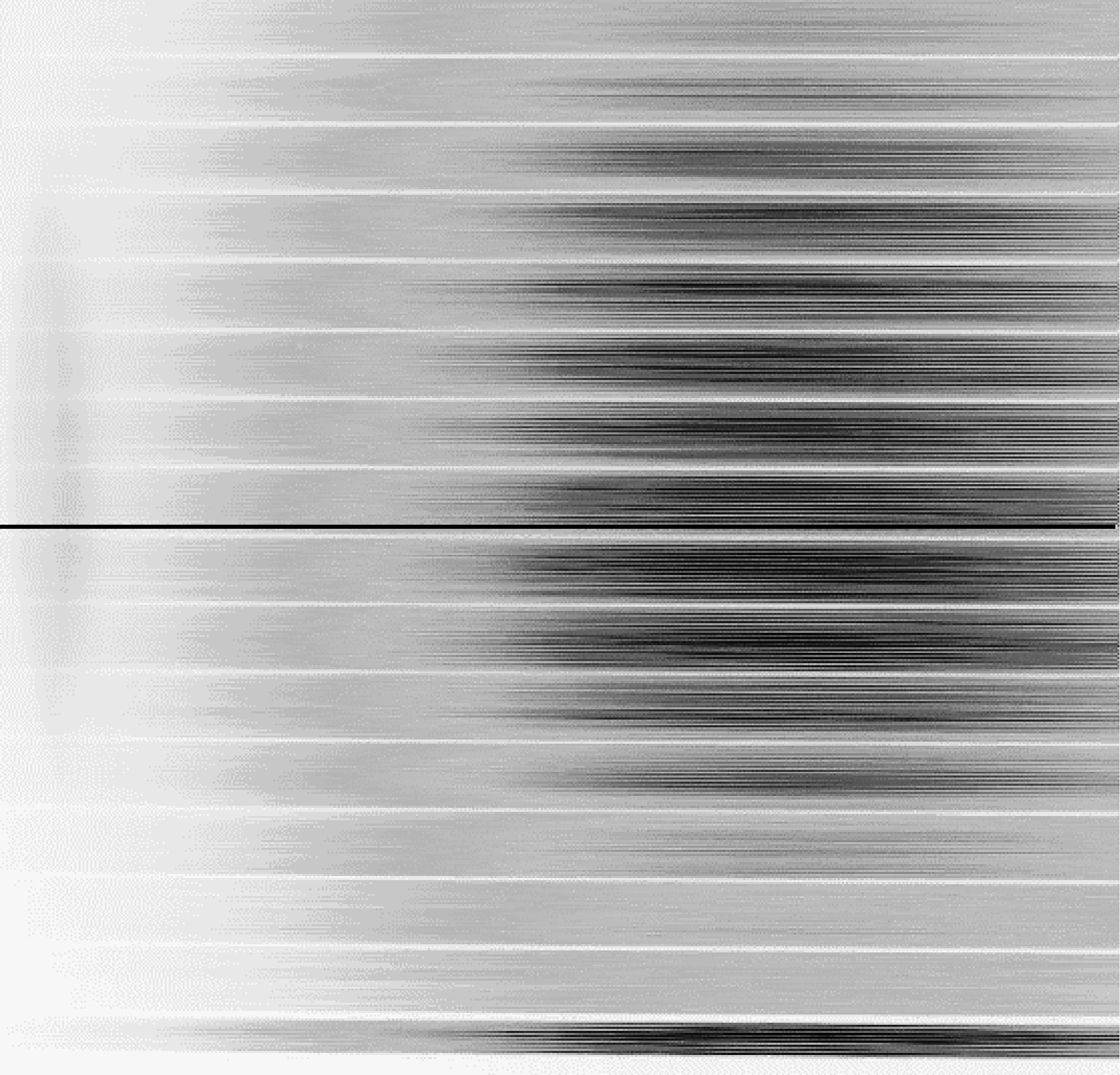}
\end{minipage}
\hspace{5mm}
\begin{minipage}[b]{5cm}
\begin{minipage}[b]{5cm}
\includegraphics[width=5cm,angle=0]{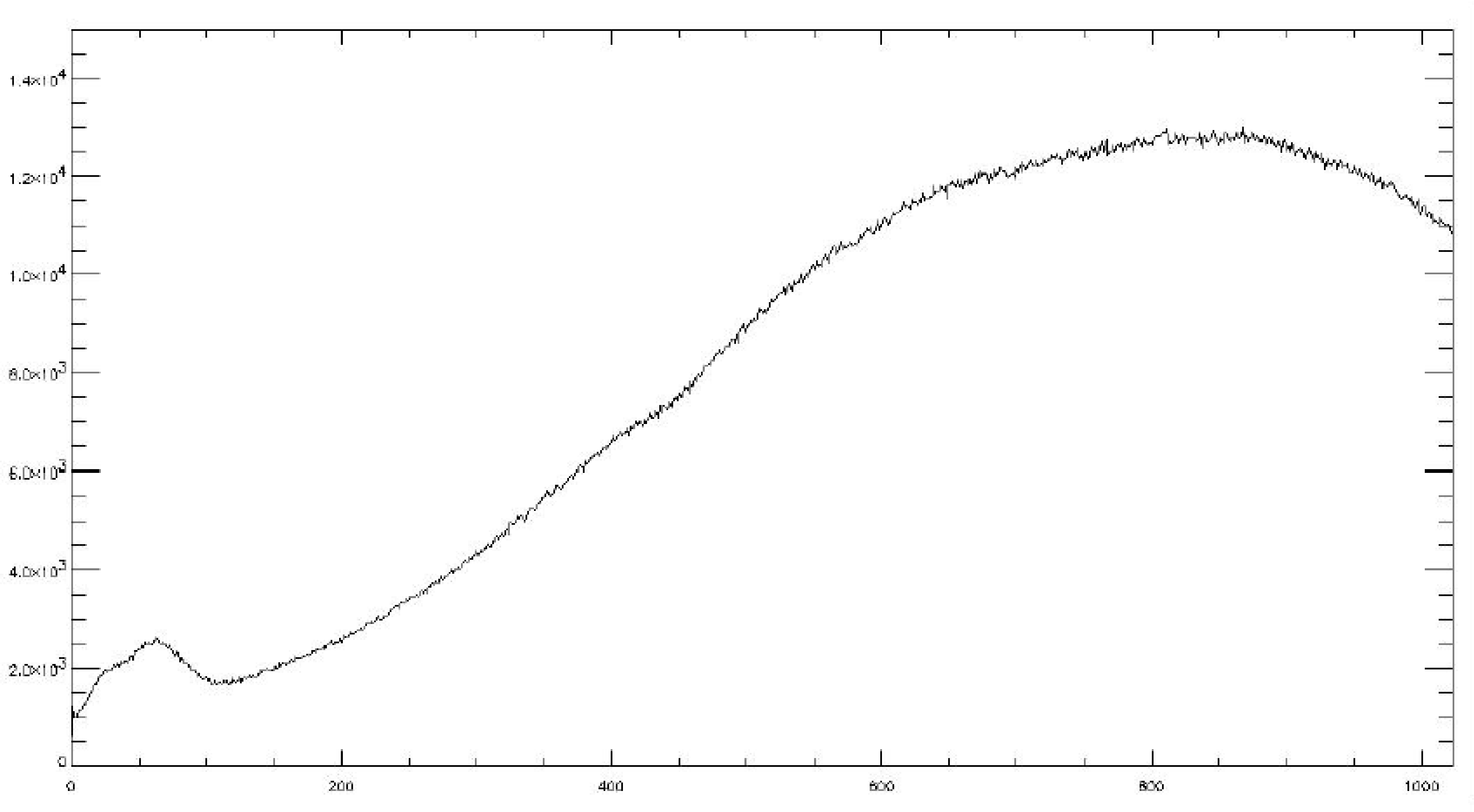}\\
\end{minipage}
\begin{minipage}[b]{5cm}
\includegraphics[width=5cm,angle=0]{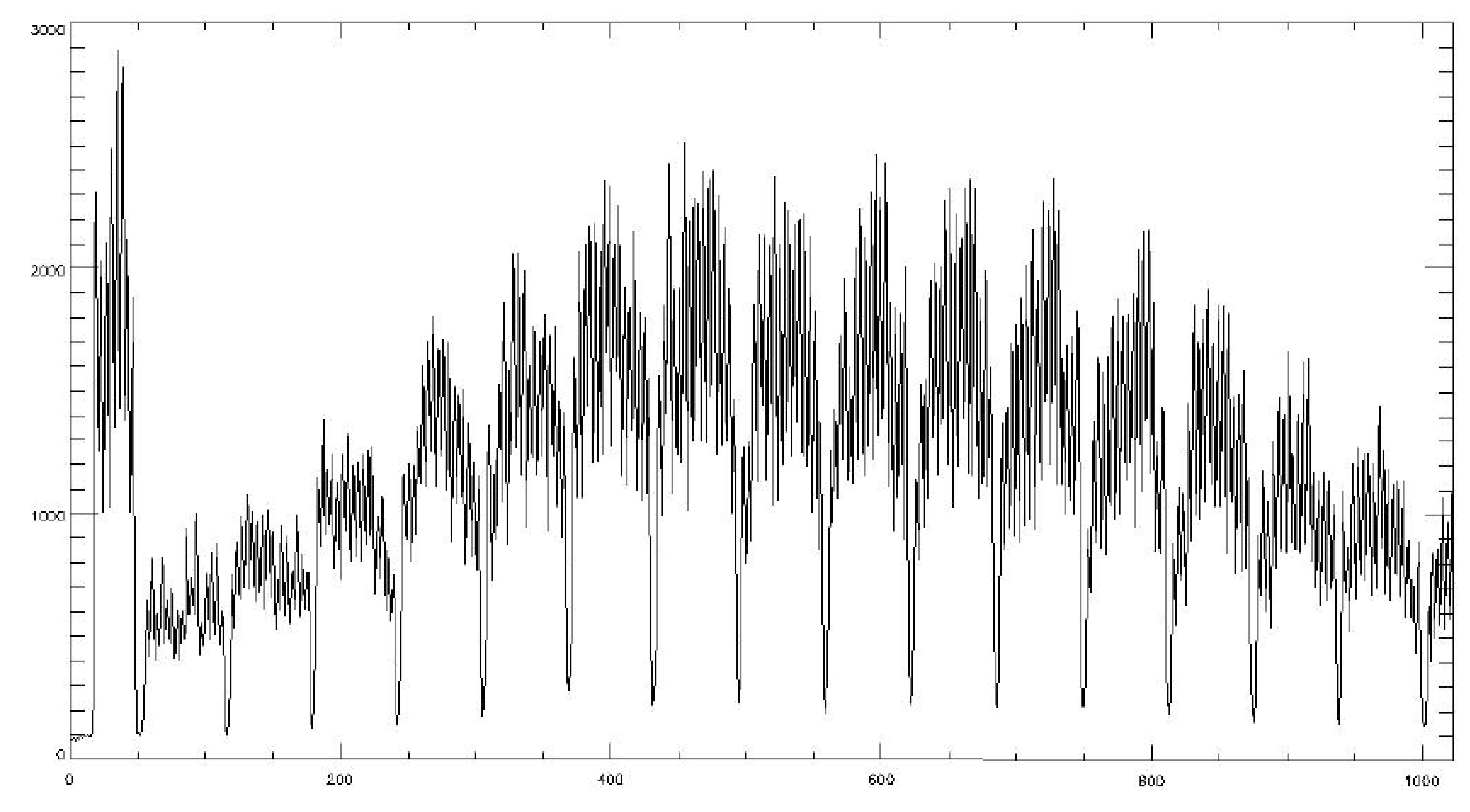}
\end{minipage}
\end{minipage}
\caption{Extraction of IFU spectra from a raw CCD frame.}
\label{EXTRACTION}
\end{figure}

\subsubsection*{Data Formats}

\begin{figure}[h!]
\includegraphics[width=12cm,angle=0]{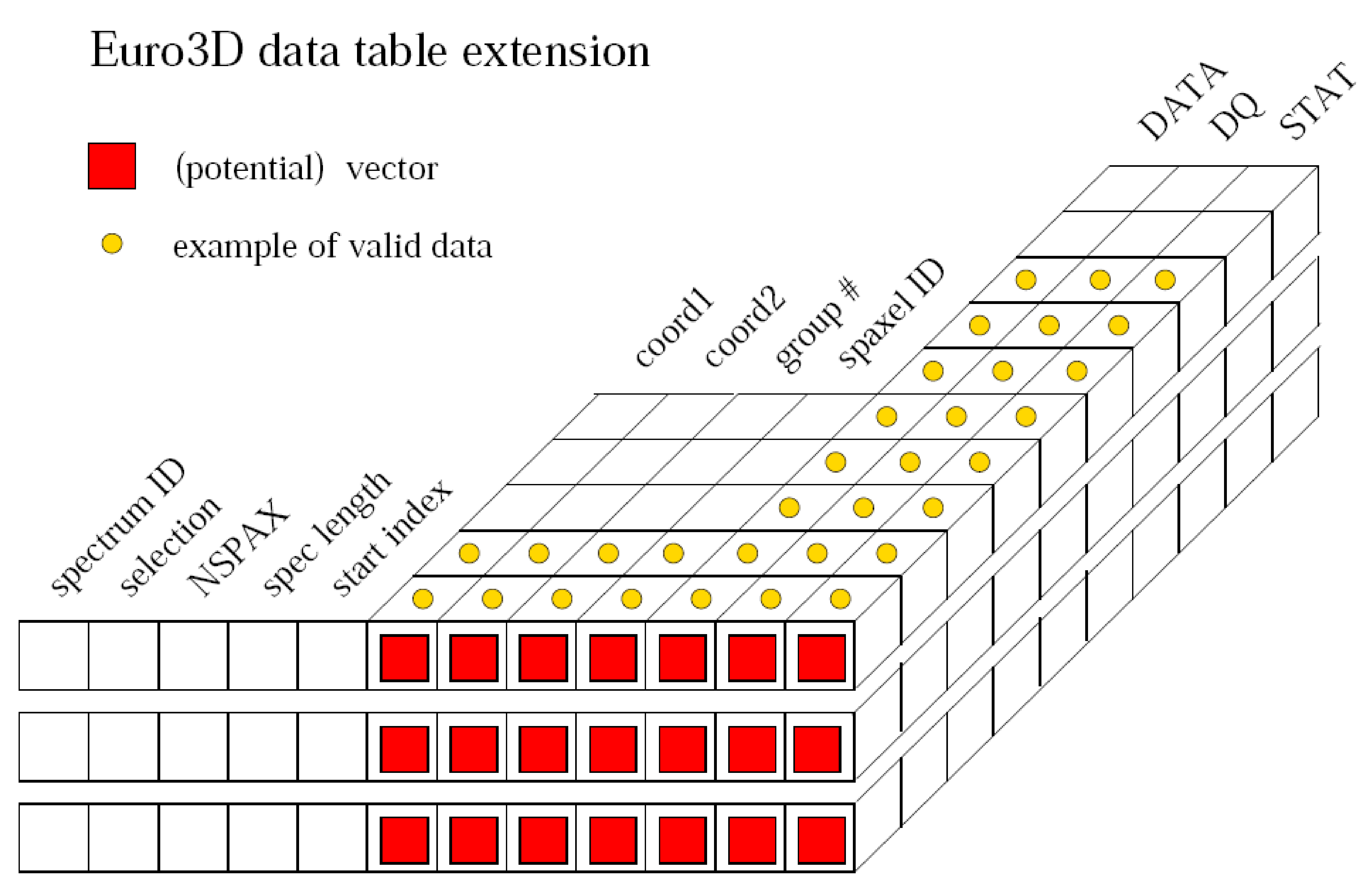}
\caption{The Euro3D data format \citep[from][]{Ki:04}.}
\label{Euro3DDATAFORMAT}
\end{figure}

The generic data product from IFS is a set of spectra, which are associated
with a corresponding set of spaxel positions. The spaxel coordinate system
may or may not be ortho-normal, but in the most general case it is not.
It is only through the process of interpolation in the spatial coordinate
system that arbitrary IFU geometries are converted to ortho-normal, i.e.
a datacube compatible form. Interpolation, however, inevitably incurs loss of
information. Therefore the Euro3D consortium has introduced 
a special data format for transportation of reduced 3D data which is different
from the seemingly simple application of the standard FITS NAXIS=3 format,
which is suitable e.g.\ for radio astronomy \citep{We:81}. 
The Euro3D data format \citep{Ki:04} avoids this latter 
step of interpolation and assumes only
that the basic steps of data reduction have been applied to remove the
instrumental signature, but else presenting the data as a set of spectra
with corresponding positions on the sky. This approach leaves spatial
interpolation and the creation of maps the process of data visualization and 
analysis, i.e.\ under control of the user. Figure~\ref{Euro3DDATAFORMAT}
illustrates the spaxel-oriented approach of the Euro3D FITS data format.

\section{IFS Visualization}

The visualization of IFS data is confronted with two fundamental requirements:
the \emph{inspection} of data for the purpose of monitoring data quality and 
correcting defects,
and the \emph{analysis} of data, i.e. the derivation of physically
meaningful quantities. Ideally, a visualization tool should support both.
While the former issue requires to preferentially look into
\emph{basic elements} of a dataset --
for example to identify detector faults that mimic a signal, to check the
correctness of the various calibration steps (bias subtraction, flat-field
and wavelength calibration, extraction of spectra) and so forth --
the latter addresses several 
possible projections of the data set. For example, the user is often
interested in obtaining a \emph{map} at one or several wavelengths over the FoV,
corresponding, for example, to emission lines of an extended gaseous object --
in order 
to create line ratio maps, from which one can derive quantities like electron 
temperature, density, dust extinction etc. On the other hand, for a subset of 
spaxels that cover a peculiar object, one might wish to co-add the flux within
a user-defined aperture, and plot the resulting \emph{spectrum}. 

\begin{figure}[t!]
\includegraphics[width=\textwidth,angle=0]{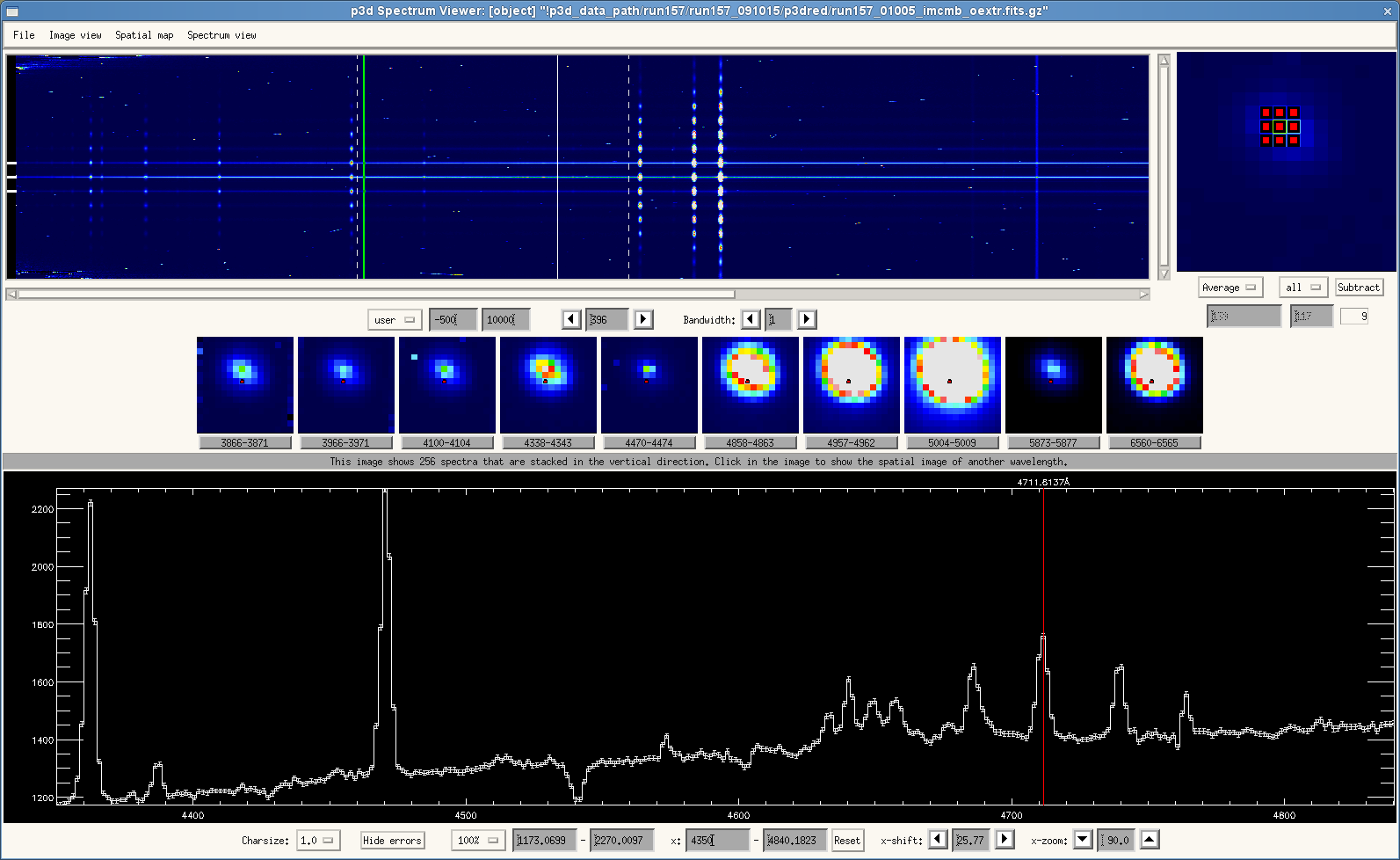}
\caption{A screenshot of the p3d spectrum viewer for a planetary nebula. 
The different regions of the tool show the spectrum image (in the top-left
region, cf.\ Fig.~\ref{fig:p3dsi}), a spatial map of a selected wavelength 
(in the top-right region, cf.\ Fig.~\ref{fig:p3dsj}), a set of ten stored 
spatial maps for ten wavelengths and a status line (middle region),
and an average 
plot of the nine spectra that are selected in the spatial map 
(in the bottom region, cf.\ Fig.~\ref{fig:p3dsp}).}
\label{fig:p3d}
\end{figure}

\begin{figure}[h!]
\includegraphics[width=\textwidth,angle=0]{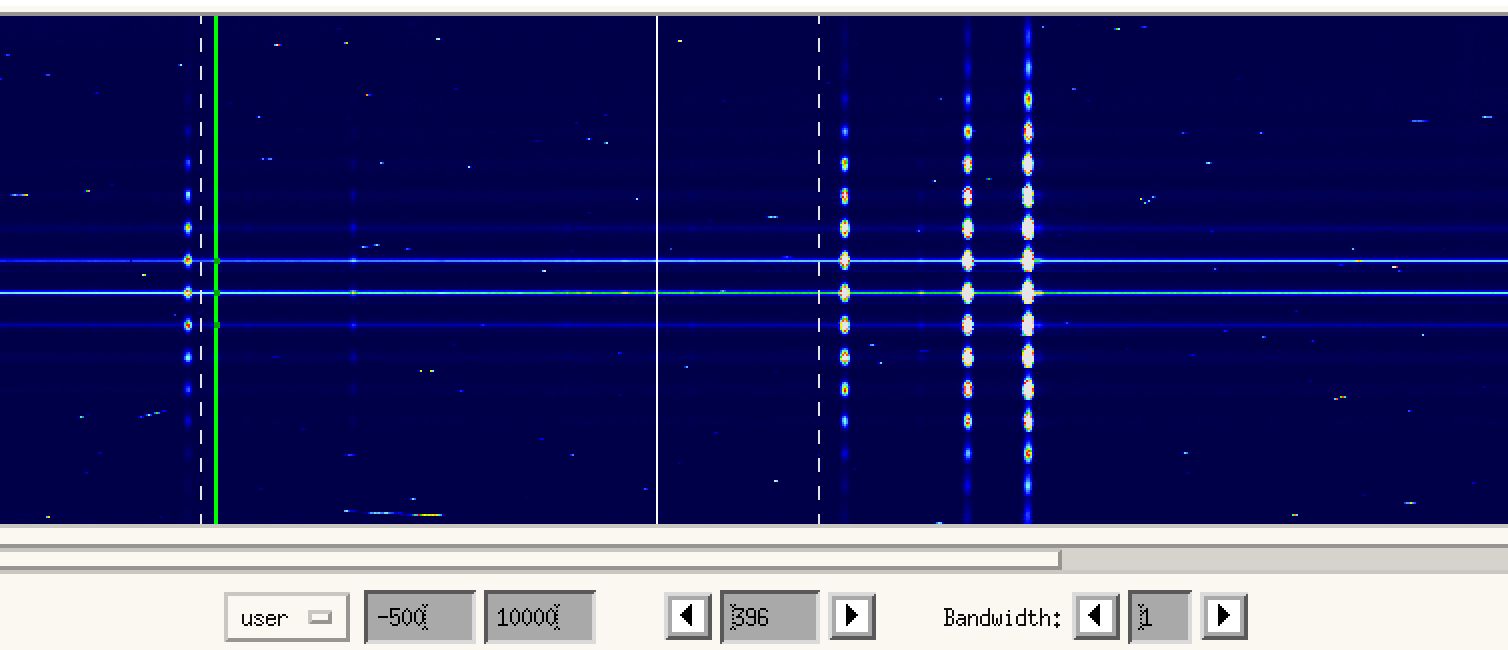}
\caption{Each line in this image represents a spectrum. The 
brighter features, which are ordered vertically, represent 
emission lines of H$\gamma$, H$\beta$, and two forbidden 
lines of $O^{2+}$ ($[\mbox{O}\textsc{iii}]\,\lambda\lambda4959,5007$). 
The brighter horizontal lines contain the central star continuum, 
and the randomly placed features are residuals of cosmic rays. 
The controls at the bottom allow to set the color cut levels, 
select the wavelength bin that is used to show the spatial map 
(the green vertical bar), and to define the width of the green bar.}
\label{fig:p3dsi}
\end{figure}

\begin{figure}[h!]
\begin{minipage}[h]{\textwidth}
\centering
\includegraphics[width=4.9cm,angle=0]{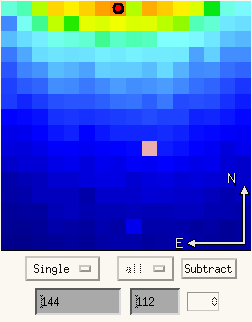}
\hspace{3mm}%
\includegraphics[width=4.9cm,angle=0]{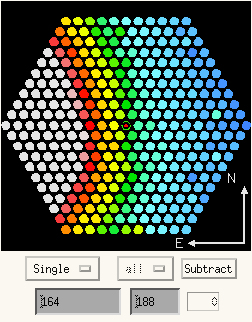}
\end{minipage}
\caption{The spatial map presents an intensity image as it is seen 
on the sky. The left-hand (right-hand) side image shows one)
selected spatial element, among all 256 (331) square-shaped (circular) 
spatial elements of the PMAS/LARR (PMAS/PPAK) IFU. The orientation of 
the IFU is also indicated, north is up and east to the left.
The controls at the bottom show the spectrum id, among other things.}
\label{fig:p3dsj}
\end{figure}

\begin{figure}[t!]
\includegraphics[width=\textwidth,angle=0]{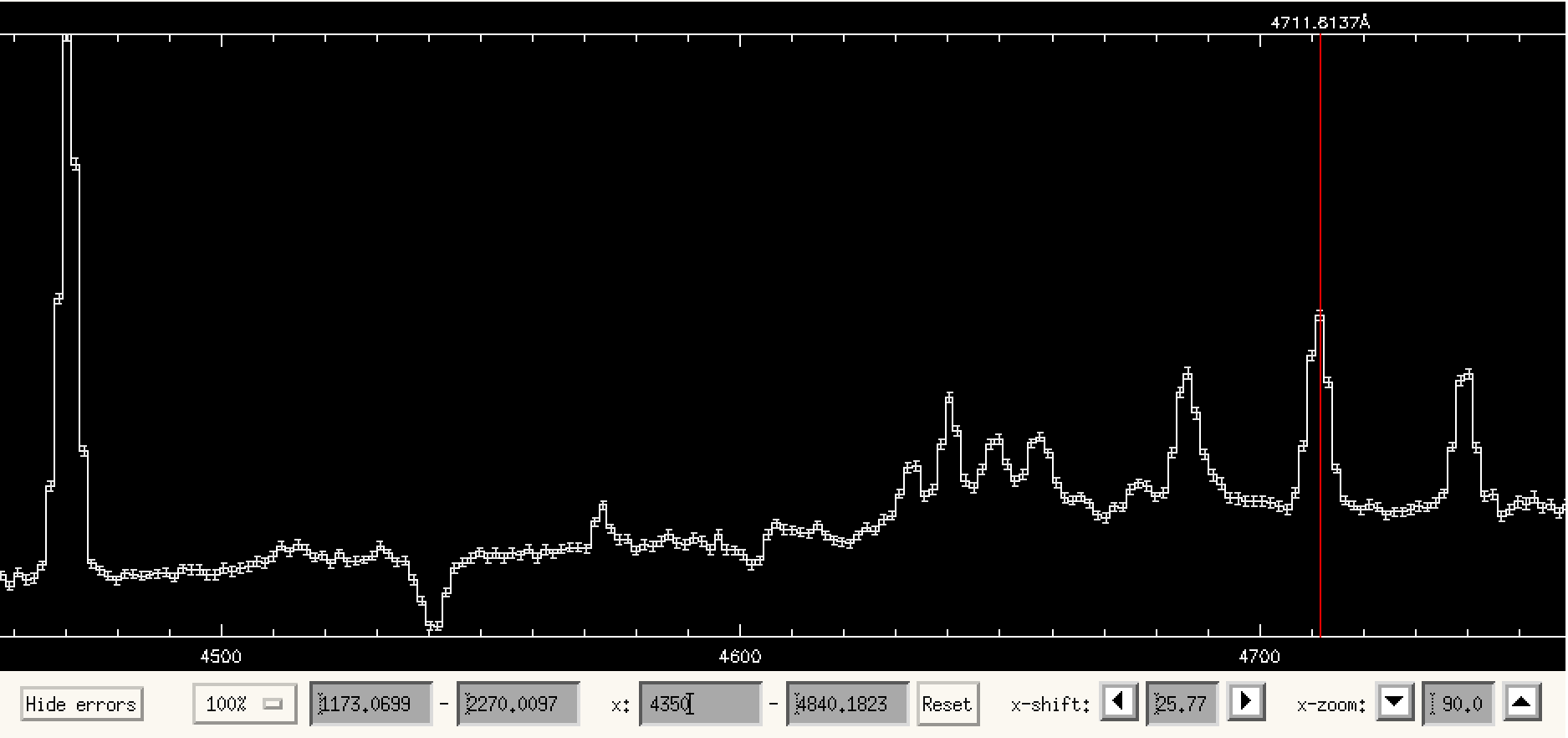}
\caption{This image shows a section of the full wavelength range of the
data for the selected spectrum, or, alternatively, of a set of averaged
or summed spectra.
Errors are shown with bars. The emission line at the cursor location, 
which is indicated with a red vertical line here (and simultaneously with
a white vertical 
line in the spectrum image, Fig.~\ref{fig:p3dsi}) comes from a forbidden 
line of Ar$^{3+}$. The controls at the bottom allow a quick change of 
the properties of the shown spectrum.}
\label{fig:p3dsp}
\end{figure}

The p3d software \citep{Sa:10} is a versatile data reduction package for
optical fiber based 3D instruments, which contains a visualization tool that
supports a variety of these needs. Its capabilities are illustrated in
Figs.~\ref{fig:p3d}--\ref{fig:p3dsp}. 

p3d is a free distribution that is licensed under GPLv3.
It is available from the project website at {\makro}.
Although p3d is coded using the
Interactive Data Language (IDL) it
can be used with full functionality without an IDL license.

\clearpage

\end{document}